\documentclass[a4paper,10pt,twocolumn]{article}

\usepackage[english]{babel}
\usepackage[utf8]{inputenc}
\usepackage[T1]{fontenc}

\usepackage[top=1.5cm, left=1.5cm, right=1.5cm, bottom=1.5cm]{geometry}

\renewenvironment{abstract}{\bf\small {\em\ Abstract---}}{}

\usepackage{amsfonts,amssymb,amsmath,amsthm}
\usepackage{subfigure}
\usepackage{graphicx}
\usepackage[footnotesize]{caption}

\usepackage{amsfonts,amssymb,amsmath,amsthm}
\usepackage{caption}
\DeclareMathOperator*{\argmin}{arg\,min}

\title{Data-driven cortical clustering to provide a family of plausible
solutions to M/EEG inverse problem}

\author{Kostiantyn Maksymenko$^{1,2}$, Maureen Clerc$^{1,2}$ and Theodore Papadopoulo$^{1,2}$.\\
  \footnotesize $^1$University of Côte d'Azur, France.\ $^2$INRIA Sophia Antipolis, France. } \date{\empty} 

\begin{document}

\maketitle

\begin{abstract} 
The M/EEG inverse problem is ill-posed. Thus additional hypotheses are needed to constrain the solution space. In this work, we consider that brain activity which generates an M/EEG signal is
a connected cortical region. We study the case when only one region is active at once.
We show that even in this simple case several configurations can explain the data. As opposed
to methods based on convex optimization which are forced to select one possible solution, we propose an
approach which is able to find several ”good” candidates - regions which are different in term of their sizes
and/or positions but fit the data with similar accuracy.
\end{abstract}

\section{Introduction}
\label{sec:introduction}
Human magneto-/electroencephalographic (M/EEG) \cite{meeg} source localization aims to reconstruct
the current source distribution in the brain from one or more maps of potential differences measured noninvasively from electrodes on the scalp surface (EEG), or maps of magnetic fields measured by magnetometers (MEG). A common approach is to represent the cortex as a finite set of current dipole sources. The M/EEG signal generated by such a source with a unit amplitude is called lead field and computed as a solution of M/EEG forward problem. Thus any measured signal can be modeled as a linear combination of lead fields associated to each dipole source
\begin{equation}
\label{eq:model}
\mathbf{y} = L \cdot \mathbf{x} + N
\end{equation}
where $\mathbf{y} \in \mathbb{R}^n$ is the signal measured by $n$ sensors, $L$ is the $n \times m$ lead field matrix whose columns represent lead fields of $m$ sources, $\mathbf{x} \in \mathbb{R}^m$ is a vector of sources amplitudes and $N \in \mathbb{R}^n$ is additive noise. In this work we do not take into account the time component of the signal. The inverse problem aims to find $\mathbf{x}$ knowing $\mathbf{y}$ and $L$.

We usually consider all vertices of a cortical mesh as possible source positions which results in $m >> n$. Thus recovering $\mathbf{x}$ from $\mathbf{y}$ in (\ref{eq:model}) is an ill-posed problem. In this work, we consider that brain activity which generates an M/EEG signal is
a connected cortical region, i.e. there is a path between every pair of vertices. We study the case when only one region is active at once and each dipole in the active region has the same amplitude, i.e $x_i = a$ if i-th source is inside active region and $x_i = 0$ otherwise.

Numerous methods were proposed to solve this problem \cite{inverse, sparse, tv}. Most source reconstruction methods are based on convex optimization and in consequence identify a single solution. But because of ill-posedness of the problem, it is highly likely that other spatially distinct source configurations can explain the data as well as the identified solution, even under such a strong constraint of being a connected region with constant activity. We propose a method based on agglomerative hierarchical clustering \cite{clustering}, whose objective is not to reconstruct one active cortical region but to find several ”good” candidates - regions which are different in term of their sizes and/or positions but fit the data with similar accuracy.
   
\section{Clustering algorithm}
\label{sec:algo}
We define a cluster $c_i$ as a set of vertices of a cortical mesh and denote its size by $|c_i|$. $L(c_i)$ denotes the lead field of a cluster, which, by construction, is a sum of  the lead fields of its dipole sources. We initialize our procedure considering each vertex as a cluster. So the vector $\mathbf{x}$ corresponds to the amplitudes of initial clusters. Their initial neighborhood is defined by a cortical mesh. Two vertices are neighbors if they share an edge on the mesh. For any pair of clusters $c_i$ and $c_j$ we define a \textbf{potential error}:
\begin{equation}
\label{eq:error}
E(i,j) = \min_a \|y - a(L(c_i) + L(c_j))\|_2 + R(i,j)
\end{equation}
where $y$ is the M/EEG data to fit, $L(c_i) + L(c_j)$ represents the lead field that we would be obtained by merging the clusters. The first term of the sum represents the data fitting error that would be obtained if the clusters are merged. $R(i,j)$ represents a regularization term which we will discuss in section \ref{sec:regularization}. We represent the neighborhood information between clusters as a function $N(i,j)= N(j,i) = 1$, if clusters $c_i$ and $c_j$ are neighbors and $0$ otherwise. $N$ is initialized based on the neighborhood of vertices.
We denote $A$ as the set of current clusters. 
 Based on this we initialize $N$ and $A$ and proceed with the following steps:
 
\textbf{Step 1.} Examine all \textbf{inter neighbors} potential error (\ref{eq:error}) and merge the clusters which minimize it: 
\[
i^*, j^* = \argmin_{i,j \in A;~N(i,j)=1}E(i,j); ~c_k = c_{i^*} \cup c_{j^*}
\]

\textbf{Step 2.} Compute lead field for new cluster: $L(c_k) = L(c_{i^*})+L(c_{j^*})$

\textbf{Step 3.} Replace two merged clusters by new cluster and update neighborhood information:
$
A = A\setminus\{c_{i^*},c_{j^*}\}\cup \{c_{k}\}
$

\[
\forall i \in A: N(i,k) = 
\begin{cases}
1, \text{ if } N(i,i^*)=1 \text{ or } N(i,j^*)=1\\
0, \text{ otherwise}
\end{cases}
\]

\textbf{Step 4.} Return to step 1 and repeat until the whole cortex is one cluster.

Merging two clusters can be seen as growing one region in the direction that locally minimizes the regularized data
fitting error. Taking into account the neighborhood information guarantees connected regions. The way we compute a lead field for new clusters constrains these regions to have constant activity, i.e. all
dipoles of active region have the same amplitude.

\begin{figure}[h]
\centering
        \includegraphics[width=0.3\textwidth]{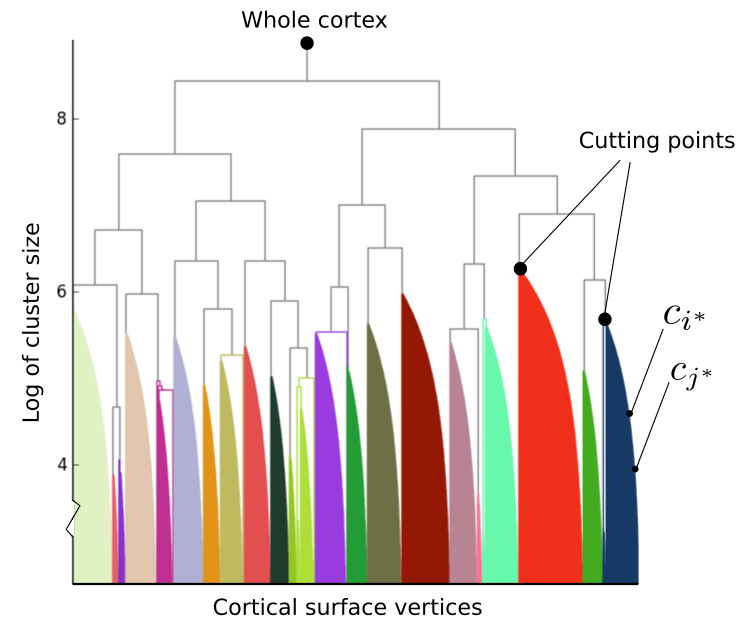}

	\caption{Example of a dendrogram that we get as a result of our clustering algorithm. The y axis represents the logarithm of clusters size for visualization purposes. We can see the particular structure of the tree being a set of smoothly growing sub-trees. Different colors represent extracted growing regions.}
	\label{fig:dendro}
\end{figure}

In the end, we obtain a dendrogram, which can be cut into a set of spatially separated growing cortical regions (Figure \ref{fig:dendro}). We based the cutting criteria on the "speed" of clusters merging. Let $c_k = c_{i} \cup c_{j}$ and $|c_{i}| \geq |c_{j}|$. $c_{i}$ is a cutting point if $|c_{j}|>s$, where $s$ is arbitrary chosen merging "speed" threshold.  
\begin{figure*}[ht]

\begin{minipage}{0.20\textwidth}
  \centering
\includegraphics[width=1\textwidth]{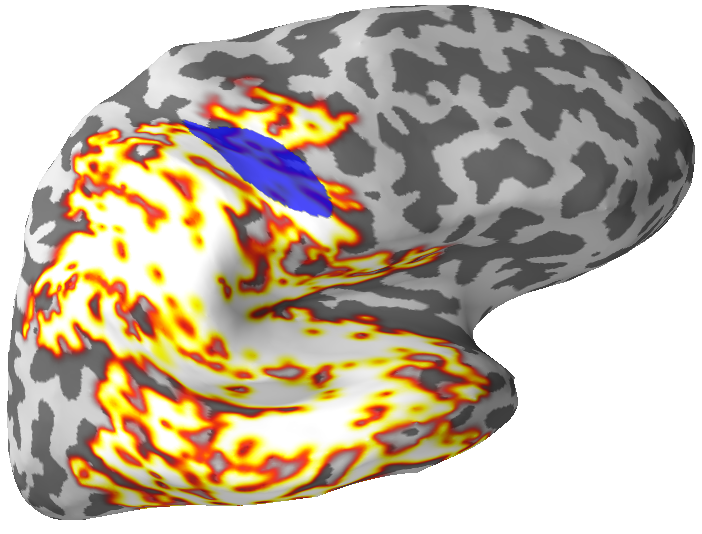}
a)
\end{minipage}%
\hfill
\begin{minipage}{0.20\textwidth}
  \centering
\includegraphics[width=1\textwidth]{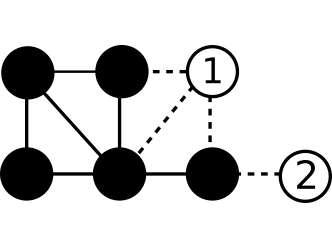}
b)
\end{minipage}%
\hfill
\begin{minipage}{0.20\textwidth}
  \centering
\includegraphics[width=1\textwidth]{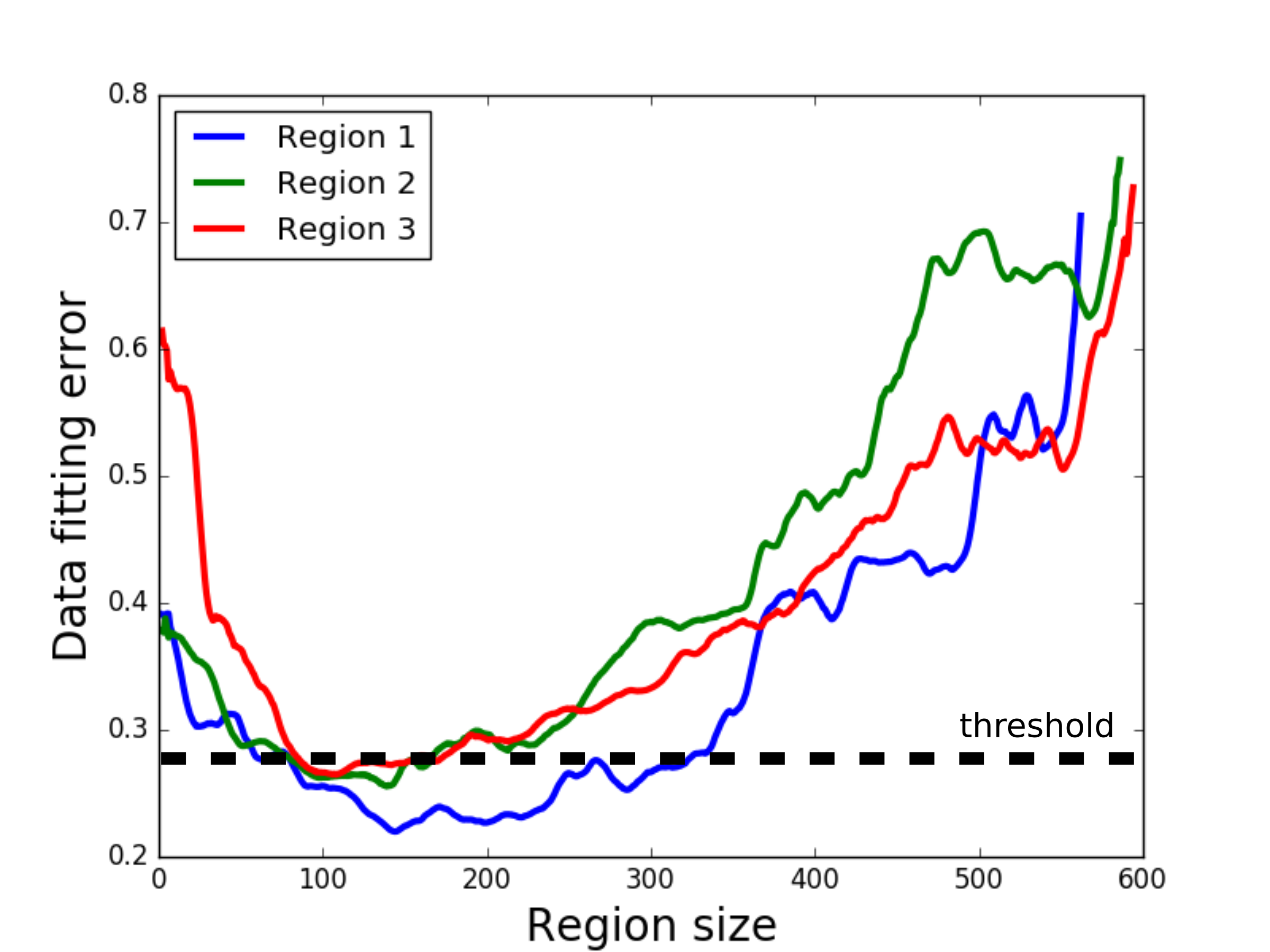}
c)
\end{minipage}%
\hfill
\begin{minipage}{0.20\textwidth}
  \centering
\includegraphics[width=1\textwidth]{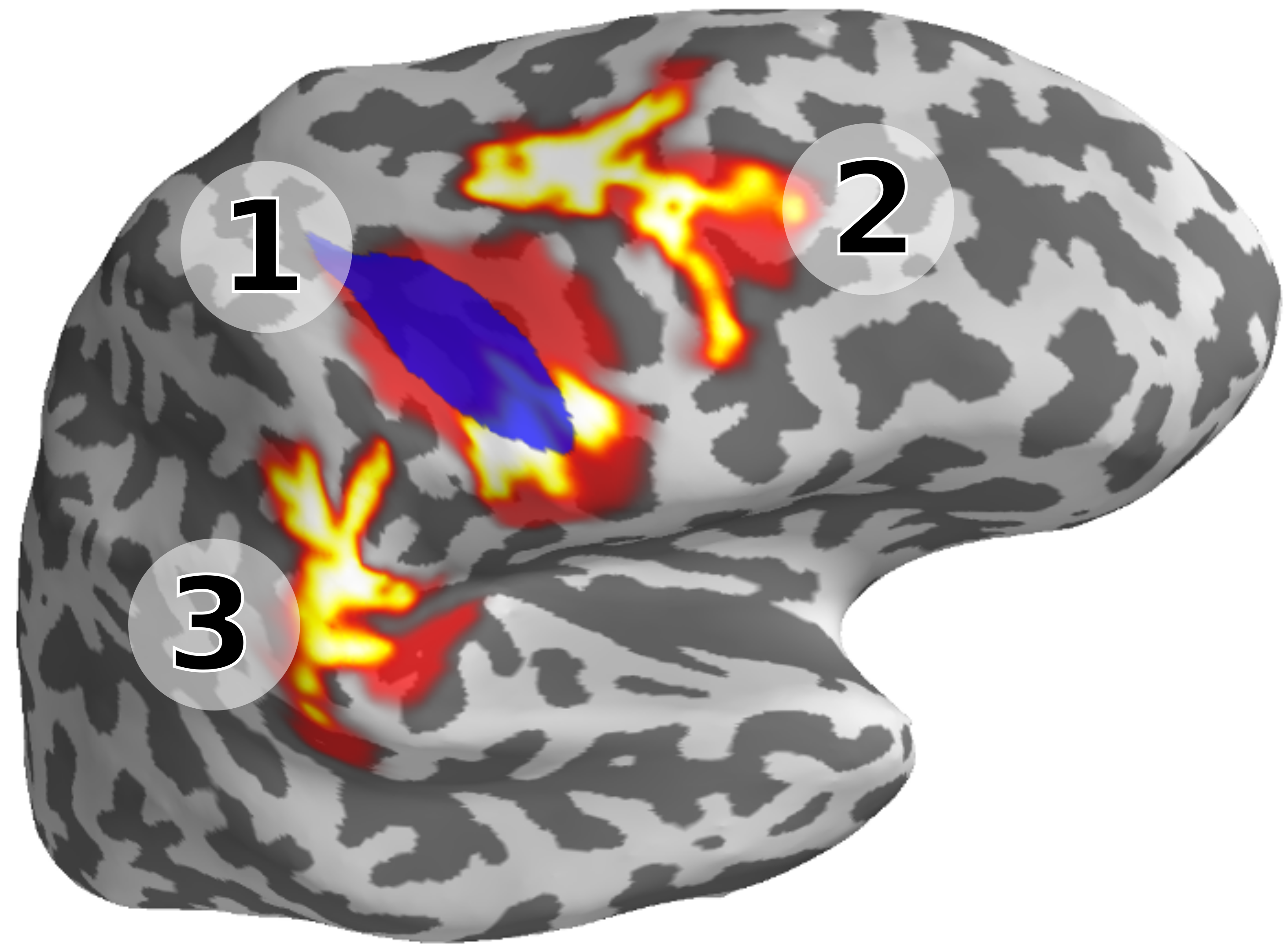}
d)
\end{minipage}

\caption{Simulation results: a) simulated region (blue) and reconstructed one (white) without regularization. b) we have two candidates (Point 1 and Point 2) to merge with the cluster of black points. For Point 1 $B(i,j) + B(j,i) = 3 + 1 = 4$ and for Point 2 it is only 2. It results in smaller regularization value (\ref{eq:regular}) for Point 1. Let us notice that merging Point 1 makes the cluster more isotropic compared to Point 2. c) Fitting error as a function of region size for top 3 reconstructed growing regions whith isotropy regularization. d) Localization of top 3 regions. Error threshold defines lower and upper bounds of their sizes (yellow and red colors resp.)} \label{fig:results}
\end{figure*}

\subsection{Regularization}
\label{sec:regularization}

Without regularization ($R(i,j) = 0$) we faced a kind of overfitting problem. The algorithm finds a way not only to reconstruct a region which has a good data fitting error, but also to keep growing the region without significant changes in error (Figure \ref{fig:results}a). We belive that it is due to merging dipoles whose lead fields cancel each other out, and by merging "blind" sources - dipoles with a low lead field norm. To solve this problem several regularization approaches can be used: penalize the size of the regions, i.e. do not allow regions to have a big size; introduce a default cortical atlas (e.g. Desikan-Killiany atlas \cite{atlas}) and allow regions to grow only inside its parcels. In this paper we introduce an approach to regularize regions' isotropy, i.e. to penalize regions with "sharp" borders and holes. 
  
For two clusters $c_i$ and $c_j$ let us define a value $B(i,j)$ as a number of vertices in $c_i$ which have at least one neighbor in $c_j$. We propose following regularization term:

\begin{equation}
\label{eq:regular}
R(i,j) = \lambda \cdot (|c_i|+|c_j|)^2\cdot \dfrac{\min(|c_i|,|c_j|)}{B(i,j) + B(j,i)}
\end{equation}

The ratio term in (\ref{eq:regular}) measures the relative length of the border between clusters $c_i$ and $c_j$. According to this measure Point 1 in Figure \ref{fig:results}b) is more favorable than Point 2 to be merged with the cluster of black points, because it has a longer "merging border". Minimizing this measure favors regions with smooth borders.

 The term $\lambda \cdot (|c_i|+|c_j|)^2$ controls the importance of regularization. Being a quadratic function of a cluster size, regularization lets regions grow freely at the beginning and starts to be important for relatively big regions. The hyperparameter $\lambda$ defines how fast regularization becomes important.

\section{Results}
\label{sec:second-section}

We used the "Sample" subject from MNE-python software data set \cite{mne} and computed its MEG forward problem. We represented source space as a cortical mesh with about 10000 vertices per hemisphere. Dipole orientations were fixed to be orthogonal to the cortical surface. We simulated one active region with additive noise and applied our reconstruction algorithm. As an output of the algorithm we obtained a set of growing regions and a data fitting error changing with respect to the region size (Figure \ref{fig:results}c). With arbitrary chosen error threshold we can select the best regions as well as their lower and upper bound sizes (Figure \ref{fig:results}d). As we can see, three spatially separated regions can explain the data with a high accuracy.

\section{Conclusion}
\label{sec:conclusion}
Our results show that even with a constrained model having only one active region with constant activity, we generally cannot find a unique data-driven solution which is significantly better than others. 

The advantage of our method is the concept of spatially separated growing cortical regions.  Compared to the methods based on convex optimization, which return a unique source configuration explaining the data, this concept provides more information about the inverse solution. It provides a relatively small number of candidate regions and estimates their size bounds.

The main directions for future work are to investigate the regularization term and the choice of hyper-parameter; to extend the method for the case when several regions are active at the same time to our clustering approach (for example, adapting MUSIC algorithm \cite{music}) and to perform a multimodal approach (simultaneous EEG and MEG acquisition) to decrease spatial uncertainty of inverse solution.

\textbf{Acknowledgment:} \textit{This work has received funding from the European Research Council (ERC) under the European Union's Horizon 2020 research and innovation program (ERC Advanced Grant agreement No 694665 : CoBCoM - Computational Brain Connectivity Mapping).}


\end{document}